CAMBRIDGE
UNIVERSITY PRESS

CrossMark

# The Local versus the Global in the history of relativity: The case of Belgium

Sjang L. ten Hagen

Vossius Center for the History of Humanities and Sciences, Institute for Theoretical Physics, University of Amsterdam and
Current affiliation: Institute for History, Leiden University
Email: s.l.tenhagen@gmail.com

**Argument**

This article contributes to a global history of relativity, by exploring how Einstein's theory was appropriated in Belgium. This may sound like a contradiction in terms, yet the early-twentieth-century Belgian context, because of its cultural diversity and reflectiveness of global conditions (the principal example being the First World War), proves well-suited to expose transnational flows and patterns in the global history of relativity. The attempts of Belgian physicist Théophile de Donder to contribute to relativity physics during the 1910s and 1920s illustrate the role of the war in shaping the transnational networks through which relativity circulated. The local attitudes of conservative Belgian Catholic scientists and philosophers, who denied that relativity was philosophically significant, exemplify a global pattern: while critics of relativity feared to become marginalized by the scientific, political, and cultural revolutions that Einstein and his theory were taken to represent, supporters sympathized with these revolutions.

## Introduction

In their 2014 *The History Manifesto*, historians Jo Guldi and David Armitage argue that the focus of historical scholarship has become increasingly narrow. They draw attention to the negative consequences of this development, claiming that professional historians, by spending their time on crafting micro-histories, are losing their ability to provide long-term perspectives, and are isolating themselves from philosophy and society (Guldi and Armitage 2014). Although some historians of science have tried to outline a program for global and long-term historiography in recent years (Raj 2013; Secord 2004), their field has particularly suffered from such a narrow focus (Galison 2008; Kuukkanen 2012). According to the editors of a recent issue on "Transnational History of Science" in *BJHS*, "thick description" has brought historians of science much insight, but also produced new challenges: "the effort to produce a sociologically inspired history of science has led to a more realistic description of its past and current legacy, but has, however, downplayed its international dimension" (Turchetti et al. 2012, 324). This is a problem, because knowledge is not bound to local contexts; it moves through national and disciplinary boundaries (Krige 2019a; Bod et al. 2019).

Such problems are manifest in the vast literature on the reception of relativity. The theory of relativity and its main protagonist, Albert Einstein, acquired world fame just after the First World War. Historians have argued that these circumstances colored perceptions of Einstein and his theory, and they have done so by writing local, nationally-focused histories (e.g. Warwick 1992; Goenner 1993; Hu 2007; Mota et al. 2009; Fox 2018).[1] These studies have supplied valuable insights. Yet, there has been insufficient attention to how the many national cases together form an overarching transnational narrative (but see: Goldberg 1984; Glick 1987b; Brush 1999, 185; Stanley 2019). In the words of historian of science James Secord, previous scholarship on the reception of relativity "has brought out the complexity and particularity of specific national situations, but [it has] done less toward creating a global picture" (Secord 2004, 669).

---

[1] The literature is immense and expanding. The limited sample of case studies illustrates that their focus has been local. A 1987 volume edited by Thomas Glick contained case studies on France, Poland, Italy, Russia, Spain, the US, and Japan (Glick 1987a).

In this article, I examine the history of relativity in Belgium. This national focus is of course surprising given the problems that I have outlined above. Yet, the national and the transnational are not mutually exclusive: they are intrinsically linked (Walker 2012). Indeed, "national developments represent relevant levels of intellectual activity, but instead of treating them as more or less self-contained universes, it is more fruitful to consider them as embedded in transnational relations of various kinds" (Heilbron et al. 2008, 147). With the complementarity of the national and the transnational in mind, I focus on Belgium as a means towards a transnational picture.

As it turns out, the Belgian case is particularly well-suited to see how the local and the global intersect. First, this is because around 1900, the bilingual kingdom of Belgium was a meeting ground of different cultures, including Dutch-speaking or francophone Catholics, socialists, and liberals. Furthermore, the country's diverse intellectual elite was very much internationally oriented (Laqua 2013). Moreover, some of the most transformative early-twentieth-century global phenomena decisively manifested themselves in Belgium. The First World War was even staged on Belgian territory. I am also referring to macroscopic processes of socio-cultural "modernization" which infiltrated and transformed Belgian science and society.[2] Thus, one might imagine the Belgian context as a microcosm reflective of a macrocosm. This is not to say that whatever dynamics surrounding Einstein and his theory in Belgium are automatically applicable to other national contexts, but rather that the Belgian context, because of its cultural diversity and attachment to global conditions, may open up a window to transnational dynamics, and offer opportunities for comparison across national borders.

No self-contained history of relativity in Belgium exists. This is remarkable, not only because of the above explained potential of the Belgian case to shed light on global patterns and transnational relations, but also because Belgium provides a rich context in its own right.[3] The German occupation of Belgium during the First World War immobilized Belgian science (De Schaepdrijver 2013). What were the ways in which Belgian scientists were introduced to relativity and what was the impact of these circumstances? Another interesting Belgian factor concerns the relatively obscure yet locally dominant gravitational theory of mathematical physicist Théophile de Donder (1872-1957). De Donder's theory provides an excellent circumstance to study the integration of relativity in a local context, and how in that context the local and global "interpenetrate" (Krige 2019b, 13-17). How did the war influence de Donder's failed attempts to learn or contribute to general relativity? How did de Donder's peculiar approach to relativity influence the work of his students? And what was the impact of de Donder on the Belgian priest cum noted cosmologist Georges Lemaître (1894-1966)? Meanwhile, Catholic and freemasonic Belgian intellectuals debated the relation between modern science and religion. From the early 1920s onwards, the theory of relativity became a key factor in these debates.

---

[2] I follow Frans van Lunteren and Marijn Hollestelle's understanding of the term "modernity" as a period, namely the one between 1890 and 1930. I also adopt their assertion that "modernization" can best be understood as incorporating scientific-technological *and* socio-cultural trends, including: "urbanization and the increasing pace of city life, industrialization and the reliance on ever more powerful technologies, the advance of bureaucratic rationality and its concomitant functional specialization, and the rise of mass democracy and modern consumerism" (Van Lunteren and Hollestelle 2013, 505-506).

[3] It may be worthwhile to point out that Einstein had a Belgian uncle, and, from the late 1920s onwards, was befriended with Belgian Queen Elisabeth. The links between Einstein and the Belgian Royal family helped him to live secure a place to live and work, in the Belgian coastal town of De Haan, before moving to the U.S. in 1933. To understand the early appropriation of Einstein and his work in Belgium, however, these social relations were irrelevant.

In this article, I provide a comprehensive image of how relativity was appropriated in Belgium. This sheds light on the changing roles of science in Belgian society during the turbulent first decades of the twentieth century (building on excellent recent work, e.g. Vanderstraeten 2018; Onghena 2011; Wils 2005; Tollebeek et al. 2003). Ultimately, I address how Belgian perspectives on relativity intertwined with those in other countries in order to identify transnational networks and global patterns in the history of relativity.

In what follows, I use the terms "appropriation" and "co-creation" rather than the commonplace "reception" to denote the historical mechanisms by which relativity became a Belgian (and a global) phenomenon. This is because "reception" fails to capture the mutability of knowledge, such as relativity, that moves across national boundaries. Historians of science have pointed out that the transfer of knowledge (including theories, concepts, and methods) between different local contexts involves processes of appropriation, during which the knowledge that is being moved transforms (Kaiser 2005, 16-23; Gavroglu et al. 2008, 159-161; Raj 2013, 339-340; Bod et al. 2019, 492-493). The passive term "reception" does not account for this flexibility. Furthermore, the term imposes an asymmetrical relationship between sending and "receiving" parties. However, relativity turned from something local into something global not only by Einstein's efforts, but also by the active efforts from many other actors who were situated in a variety of local contexts, including Belgium. These instances of appropriation, or "co-creation" (cf. Krige 2012 on "co-production"), gave rise to a multitude of local versions of and perspectives on relativity (Renn 2007).

## How the war shaped the circulation and public perception of relativity in Belgium

Relativity was closely tied to theoretical physics. As Richard Staley has argued, the theory of relativity "came to represent a defining achievement of the new subdiscipline of theoretical physics," both in public and scientific spheres (Staley 1998, 265). Therefore, to understand how relativity was appropriated in Belgium, it is important to be informed about the state of Belgian theoretical physics at the beginning of the twentieth century.

The roots of theoretical physics lie in Germany, where a synthesis between experimental and mathematical approaches to the study of natural phenomena had developed over the course of the nineteenth century (Jungnickel and McCormmach 2017). Well before the turn of the twentieth century, the new and hybrid theoretical physics (sometimes still referred to as mathematical physics) gained a foothold in other European countries as well, including Belgium's neighboring countries. Hendrik Lorentz was appointed professor of theoretical physics in Leiden in 1878, and Henri Poincaré accepted a chair in mathematical physics at the Sorbonne in 1886.

In Belgium, however, theoretical physics initially failed to institutionalize. Remarkably, the first Belgian chair in theoretical physics in Brussels in 1912 was not offered to a Belgian but to a Frenchman: Émile Henriot. A year earlier, De Donder had been appointed as the first Belgian professor of mathematical physics at the Université Libre de Bruxelles (ULB). At the other three Belgian universities (in Ghent, Louvain, and Liège), chairs were only created in the 1930s (Vanpaemel 2001). Belgian priorities lay in experimental rather than theoretical (or mathematical) physics, and first and foremost in engineering. At the beginning of the twentieth century, Belgian scientists and industrialists together had sought to integrate science and industry in order to create direct social benefits; the idea was that science could forge economic dividends by means of technological applications (Onghena 2011, 290). In line with these ideals, it was considered the main duty of Belgian physicists to train future generations of engineers. An internationally oriented academic career was, at best, regarded as of secondary importance.

Tellingly, the only Belgian participant in the first editions of the prominent Solvay Councils for physics (organized in Brussels in 1911 and 1913) was an engineer.[4] According to Léon Rosenfeld, who later studied under Niels Bohr and would spend the larger part of his career outside Belgium, the condition at the University of Liège, where he studied in the 1920s, was such that "in mathematics, physics and chemistry, the upper 'layer' of the students were the future engineers. Those who were studying pure science . . . were regarded as the scum."[5] Given the above, it is understandable that De Donder wrote to his Dutch colleague Lorentz in March 1914 that "the study of mathematical physics is much behind and negligible" in Belgium.[6] Soon, Belgian physics would come to lag even further behind.

---

[4]The engineer was Robert Goldschmidt. From 1924 onwards, De Donder took part as well. More Belgian physicists, such as Léon Rosenfeld and Charles Manneback, followed in the 1930s (Marage and Wallenborn 1999, 215-217).

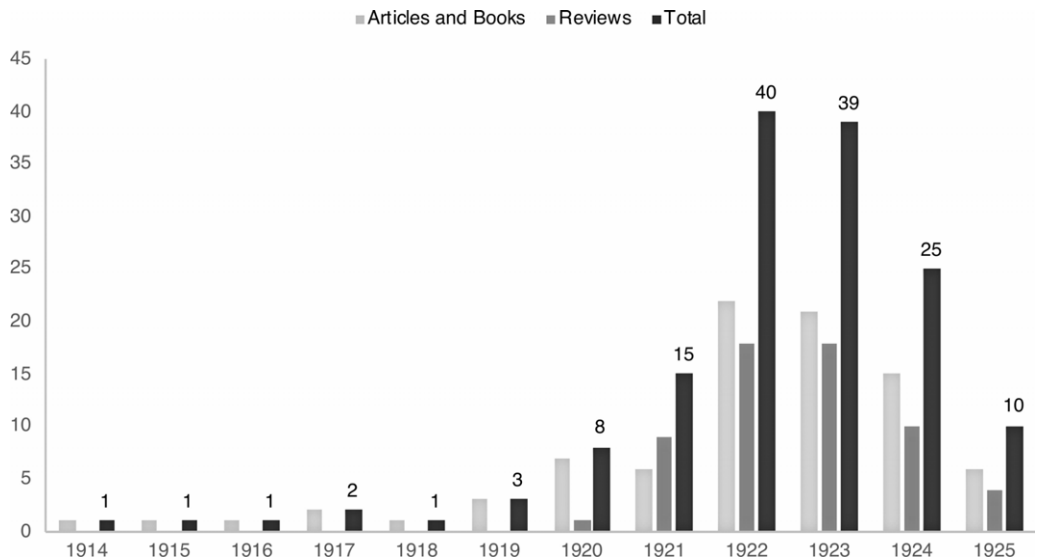

**Figure 1.** Belgian academic output on relativity physics 1914-1925 (nothing was published before 1914). Consulted sources are Lecat (1924), and J.C. Poggendorff's 5th and 6th volume of the *Biographisch-Literarisches Handworterbuch der exakten Naturwissenschaften*. Numbers include both publications on relativity by Belgians (in Belgium and abroad) and publications by Belgian editors (by Belgians and non-Belgians).

The German army invaded Belgium in the summer of 1914. On September 20, it set fire to the old university library of Louvain. During the four years of occupation that followed, Belgian universities and scientific societies were closed. Many Belgian physicists worked in exile or depended on communicating with colleagues abroad in order to remain informed about the developments in their fields. As a result, Belgian science, which had already been marginal, suffered a further severe setback. Furthermore, the technological results attained through the emphasis on applied science during the war confirmed the conviction of many Belgians that a strong bond between science and industry brought prosperity and progress (Onghena 2011, 282-290). Besides its isolating effect, the war thus reinforced the Belgian preference for the kind of physics that was not immediately geared towards industrial or economic benefit.

These circumstances are part of an explanation of the initial lack of Belgian interest in the new relativity physics. While in neighboring countries Einstein's work had caught the attention of many well before the outbreak of the war, it triggered barely any scientific activity in Belgium (fig. 1). The only Belgian scholar who did pay attention to relativity was De Donder.

---

[5]Interview of Leon Rosenfeld by Thomas S. Kuhn and John L. Heilbron on 1963 July 1, Niels Bohr Library & Archives, American Institute of Physics, College Park, MD USA.

[6]De Donder to Lorentz, 22 March 1914. Unless noted otherwise, cited letters to Lorentz have been consulted at the *Noord-Hollands Archief* in Haarlem, inventory no. 364. Translations are mine.

### De Donder's priority claim and isolation

De Donder was responsible for just about the entire Belgian corpus of relativity physics during the 1910s. De Donder's principal inspiration was Henri Poincaré: As a student of Poincaré in 1901 and 1902, De Donder had been introduced to "integral invariants," and published on this mathematical formalism which he extended during the years that followed (De Donder 1901; De Donder 1904).[7] Starting in 1913, he noticed that the Poincaréan formalism could be relevant to Einstein's (still fruitless) attempts to generalize special relativity. From that point onwards, De Donder attempted to create his own relativistic theory of gravitation, and he corresponded about it with Lorentz and Einstein.

Despite the outbreak of war, which worsened his working conditions, De Donder nevertheless thought he was well on his way to making a lasting contribution to general relativity. In 1916, in an article published in the journal of the Dutch Academy of Sciences, De Donder announced that he had successfully generalized Einstein's theory of special relativity by deriving a covariant set of gravitational field equations, a result which Einstein himself had been seeking for years and had attained in 1915 (De Donder 1916). A year later, De Donder still claimed to have obtained his results independent of Einstein. "The war has prevented me from publishing them earlier," he stated in the appendix to a 1917 publication which included his results from 1915 and 1916 (De Donder [1917] 1920a, 82). Indeed, the war had strongly limited De Donder's ability to participate in the international debates in physics. Because of the shutdown of the Belgian university system, De Donder was forced to accept a temporary position at the *Athénée* of St. Gilles in Brussels, from where he tried to keep in touch with scientists abroad and to reach an audience for his results (Glansdorff et al. 1987, 23). Before 1914, De Donder had preferred to publish in French and Belgian periodicals. During the war, however, this option was no longer available. Hence, De Donder was forced to shift attention to the Netherlands.

Between 1914 and 1918, De Donder was heavily dependent on the help of his colleague Lorentz, who sent him relevant literature and facilitated the publication of De Donder's articles in Dutch journals. These enjoyed some wider circulation, due to the fact that the Dutch maintained a central position in transnational networks of relativists. After the war, De Donder thanked Lorentz for his help: "during this long War, you have constantly helped me with the publication of my works. Thanks to you, I have remained in contact with the scientific world."[8] The effect of Lorentz's assistance had only been limited, however, as the German occupation made it difficult for De Donder to maintain a correspondence across the Dutch-Belgian border. At one point, De Donder complained that "the ways of sending manuscripts due to German diplomacy [is] very slow, because of the distrust awakened by mathematical symbols."[9]

As a result of these conditions, De Donder was not well-informed about contemporary developments in relativity physics. Most importantly, he had not heard about the successful axiomatic derivation of the field equations of general relativity by the Göttingen mathematician David Hilbert in November 1915.[10] Hilbert's approach was similar to De Donder's, as both relied on a variational method. However, Hilbert finished his works nine months before De Donder, making the latter's derivation largely redundant. Therefore, the priority claim of the Belgian physicist must have seemed absurd to other relativists (Janssen and Renn 2007, 902-903).

On top of that, De Donder's derivation contained a crucial mistake. He had neglected an indispensable feature of relativity: the curvature of space-time. In a lengthy correspondence in the summer of 1917, Einstein tried to make De Donder aware of the, from Einstein's point of view, rather trivial error.[11] Nonetheless, De Donder still expected to have his priority claims recognized. After he had corresponded with Einstein, he wrote Lorentz and came to a remarkable conclusion: "From numerous exchanges of viewpoints that I had with M. Einstein, it is clear that all essential results exposed in my work are in good harmony with those of Einstein."[12] When Einstein took note of another flawed publication by De Donder in the Netherlands in 1917 (De Donder 1917), he immediately wrote to the editor, Lorentz, to demand a correction, dismissing De Donder's work as a misguided "priority claim" ("*Prioritätsanspruch*").[13]

---

[7]Integral invariants comprise integral functions that remain constant under all possible transformations. In a pre-war letter to Lorentz, De Donder addressed their potential for relativity physics (De Donder to Lorentz, 22 March 1914).

[8]De Donder to Lorentz, 2 February 1919.

[9]De Donder to Lorentz, May 1916. Exact date illegible on the original letter. Later that year, De Donder made similar remarks (De Donder to Lorentz, 25 September 1916; De Donder to Lorentz, 30 April 1916).

[10]In the late summer of 1916, De Donder stated that "I have just taken notice of the recent *Erste Mitteilung* of M. Hilbert" (De Donder to Lorentz, 22 August 1916).

A few years after the war, De Donder had corrected his derivations, and managed to generate modest interest in his work on the foundations general relativity. These accomplishments make his initial lack of success all the more remarkable. How can this contrast be explained? Crucially, there was a technical reason behind the fruitlessness of De Donder's attempts to contribute to general relativity during the war; his work was simply incorrect, and his variational approach had already been successfully executed by Hilbert. That being said, the faulty nature of De Donder's contributions cannot be seen separately from the local setting in which they were produced.

Some months after publication of his derivation of the gravitational field equations, De Donder wrote to Lorentz to ask whether his results had already been included in Einstein's latest book.[14] De Donder's expectation that Einstein would regard his contributions to general relativity as necessarily meriting a citation in his work indicates that he had no sense for how his work was viewed outside of Brussels. During the war, the French-Belgian networks that De Donder had been part of during his early career ceased to function. Yet, it would be incorrect to ascribe De Donder's initial lack of international success in relativity physics wholly to the war-time impediment of knowledge circulation. De Donder's isolation was especially apparent during the war, but it had already existed before. With or without the First World War, the Belgian academic world would not have provided a disciplinary structure suited to appropriating the new relativity physics. This is illustrated by the fact that De Donder already reached out to Lorentz before the war, while complaining about the poor level of mathematical physics in his home country.

De Donder's case underlines the need of transnational historiography to focus not only on knowledge in circulation, but also on "the social and material constraints that *impede* the circulation of knowledge" (Krige 2019b, 2-3; emphasis original). Furthermore, it emphasizes the key role of personal contact in the transnational circulation of relativity physics. According to Andrew Warwick, connections between different communities of physicists in Cambridge, Leiden, and Berlin were not established by the mere publication and diffusion of Einstein's papers. Rather, the transnational circulation of relativity theory "was an ongoing process enabled by regular correspondence, personal interaction, discussion, and collective interpretation" (Warwick 2003, 461). For example, Einstein's visit to Leiden in 1916 allowed Dutch astronomer Willem de Sitter to learn general relativity through personal interaction with its foremost expert. De Donder, in contrast, had little coaching when he came in contact with relativity, apart from his occasional correspondence with Lorentz. Furthermore, De Donder did not enjoy an institutional setting comparable to Leiden, or Göttingen, where research into mathematically advanced physical theories, particularly relativity physics, was high on the agenda (see also Stanley 2019, chap. 8). Indeed, De Donder had to grasp the Einsteinian program from printed sources, and largely on his own. This proved too much to ask.

---

[11] The 1917 correspondence between Einstein and De Donder includes nine letters, written in French and German between June 27th and August 8th (in Schulmann et al. 1998a, 303-329; trans. in Hentschel and Hentschel 1998, 224-243).

[12] De Donder to Lorentz, 22 August 1916.

[13] Einstein to Lorentz, 18 December 1917 (in Kox 2008, 493). Somewhat later, Einstein called De Donder's work "scandalously superficial," and said he was surprised that "Lorentz was taken in by it, or accepted it without closer consideration" (Einstein to Erwin Freundlich, before 17 January 1918, in Schulmann et al. 1998b, 608-610, trans. in Hentschel and Hentschel 1998, 444-445).

[14] De Donder to Lorentz, May 1916.

## Post-war perceptions of Einstein: A "German" scientist?

The case of De Donder has already exemplified how the global condition of the First World War shaped (and impeded) the appropriation of relativity physics on a local scale. Yet, not only De Donder's isolated attempts, but also the history of relativity in Belgium at large was intimately linked to the First World War. In particular, Einstein's German nationality became a major factor in post-war Belgium, both in public and scientific spheres.

The war disrupted academic networks across Europe, which had formed relatively recently (Rasmussen 1995). Around 1900, ties between Belgian and German academics were strong, both in the sciences and the humanities. But the war put these relations under pressure (Tollebeek 2008, 208-209). A letter from the Belgian experimental physicist Jules-Émile Verschaffelt to his mentor Lorentz illustrates this: A month after Louvain's university library had gone up in flames, Verschaffelt wrote to Lorentz that he found it "astonishing and disturbing that German scholars, who surely could not have felt the slightest amount of hostility against Belgium, have been influenced by the general mood in their country to such an extent that they have not spent a word of protest against the demolition of a university."[15] Verschaffelt must have become even more indignant when he heard that, instead of issuing a collective apology, a substantial group of prominent members of the German cultural elite claimed that their country had done nothing wrong. Indeed, on 4 October 1914, 93 well-respected German intellectuals (including six physics Nobel laureates) defended the legitimacy of the actions of the German army in Belgium in the infamous "Manifesto of the 93." This triggered a long-lasting conflict in the international intellectual community (Kevles 1973; Schroeder-Gudehus 1973; Wolff 2003, 341-342). After the war, scientists from the Allied countries, including Belgium, advocated a boycott of German science. Two prominent Belgians, the astronomer Georges Lecointe and the military commander Henry de Guchtenaere, strongly objected to any future rehabilitation of scientific bonds with Germany. By distributing the Manifesto, they proclaimed, "German intellectuals have prostituted science on behalf of their own interests ... scientists and intellectuals ... cannot be excused for such unscrupulousness" (Lecointe and De Guchtenaere 1919, 7).

Einstein had refused to sign the manifesto of the 93, and he was in fact one among three signatories of a counter-manifesto (Wolff 2003, 344-345). Nonetheless, post-war anti-German sentiment had a strong influence on Belgian attitudes towards Einstein and relativity. The war had split the international intellectual community into two camps. For many, it was unclear to which camp Einstein belonged, being on the one hand German, but on the other hand Swiss, Jew, and pacifist. The post-war organization of the Solvay Conferences provides an opportunity to find out how this shaped the perception of Einstein and his theory among Belgian intellectuals.

Initiated by Belgian industrial Ernest Solvay and the German scientist Walther Nernst in 1911, the Solvay Conference on physics was one of the most prominent expressions of pre-war scientific internationalism. During its first two editions, German scientists had been well-represented. Verschaffelt referred to this in the aforementioned letter to Lorentz, saying that he found it inconceivable "that those who have used Solvay's monetary aid, have not voiced any concerns on the news that Solvay has been put into prison as a hostage" (ibid.). Verschaffelt's sentiment was broadly shared, which resulted in the exclusion of German scientists from the first Solvay conferences following the war, in 1921 and 1924 (Marage and Wallenborn 1999, 21; Onghena 2011, 284). Only after extensive mediation by Lorentz, an exception was made for Einstein (who nonetheless refused both invitations, protesting the exclusion of his German colleagues).

---

[15]Verschaffelt to Lorentz, 20 September 1914.

Lorentz reminded Solvay that "a man like Einstein, the great and profound physicist, is not 'German' in the way the word is so often used today; his judgement of the events of the past years differs little from yours or mine."[16]

Some Belgian internationalists used the science of Einstein as a means to promote their own ideals. The Catholic scientist Maurice Lecat and his wife M. Lecat-Pierlot undertook the laborious project to catalogue the *entire* international literature on relativity physics. Their result, the internationally acclaimed *Bibliographie de la Relativité,* was published in 1924. The bibliography was introduced as a reaction to global opposition to relativity which, according to Lecat, had mainly stemmed from Germanophobia: "Physicists, philosophers, doctors and authors from many other intellectual domains (including mathematicians!) have written severe nonsense in order to demolish or even pulverize Einstein!" According to Lecat, their motives had been political rather than scientific: "Too often they are blinded by hatred of the '*Boche*,' how charitable and scientific is that?" (Lecat and Lecat-Pierlot 1924, vi).

One might ask whether the kind of politically motivated attacks to which Lecat referred had also been launched in Belgium. The example of the Solvay conferences showed that, even though Belgian intellectuals commonly rejected German science after the war, they were prepared to make an exception for Einstein. Furthermore, for De Donder, the German origins of relativity never formed an impediment: He never expressed his opinion on the matter, while the number of his publications on general relativity steadily grew both during and after the war. Belgian journalists, however, had been less tolerant toward Einstein than Belgian scientists. In the early 1920s, relativity was condemned in the Belgian press, for reasons similar to those addressed by Lecat. Initially, Einstein and his theory had even been completely ignored.

Throughout the Western world, an Einstein-hype emerged following the public announcement of Arthur Eddington's solar eclipse expedition results in November 1919 (Stanley 2003; Kennefick 2019, chap. 12). In the Belgian press, however, the event was barely noticed (fig. 2).[17] Before 1922, Belgian journalists barely mentioned the internationally celebrated verification of the theory. On the rare occasions that they wrote about relativity, their main concern was whether or not its original author was a German (or, in less neutral language, a "*Boche*").

In 1921, the liberal newspaper *L'Indépendance Belge* argued for a link between the "mysterious character" of the theory of relativity and its German origins. Einstein's German nationality was considered "really annoying," because "all novel notions and methods that will be established by a German will be, in their origin, German." Einstein's refusal to sign the manifesto of 93 had only slightly comforted the author of the article, because "he still is a German."[18] The Flemish nationalists of *De Schelde* disregarded Einstein for similar reasons: "We know that Einstein refused to sign the manifesto of the 93 … but he is, in the end, although Israeli, a German scholar! A *Mof,* a *Hun,* a barbarian, a *Boche*!"[19] Another journalist writing on behalf of *L'Indépendance Belge* questioned whether Einstein's nomination for the 1921 Nobel Prize was appropriate: "Einstein, we know, is German, and the wounds that Germany has made bleed, still, in a cruel way. There will be opposition."[20]

These examples demonstrate that, after the war, Belgians regarded Einstein and his theory through a political filter. Sometimes, this actually turned out to Einstein's advantage. Indeed, Einstein was occasionally praised for being an exceptional figure within a ruined German cultural climate. A year before it dispelled Einstein and his theory for being German, *L'Indépendance Belge* had portrayed him as a dissident rather than as a typical representative of German science.

---

[16]Lorentz to Solvay, 10 January 1919 (in Pelseneer 1946, 40-43). I thank Kenneth Bertrams for sending me a copy of the unpublished manuscript.

[17]Being one of the very few exceptions, the *Gazette de Charleroi* reported on an announcement by the French Academy of Science: *Gazette de Charleroi,* 11 December 1919, "Les théories d'Einstein et de Newton," 2.

[18]"Einstein," *L'Indépendance Belge,* 23 August 1921.

[19]"Wetenschappelijke Toenadering," *De Schelde,* 27 November 1921.

[20]"Einstein et le Prix Nobel," *L'Indépendance Belge,* 13 October 1921.

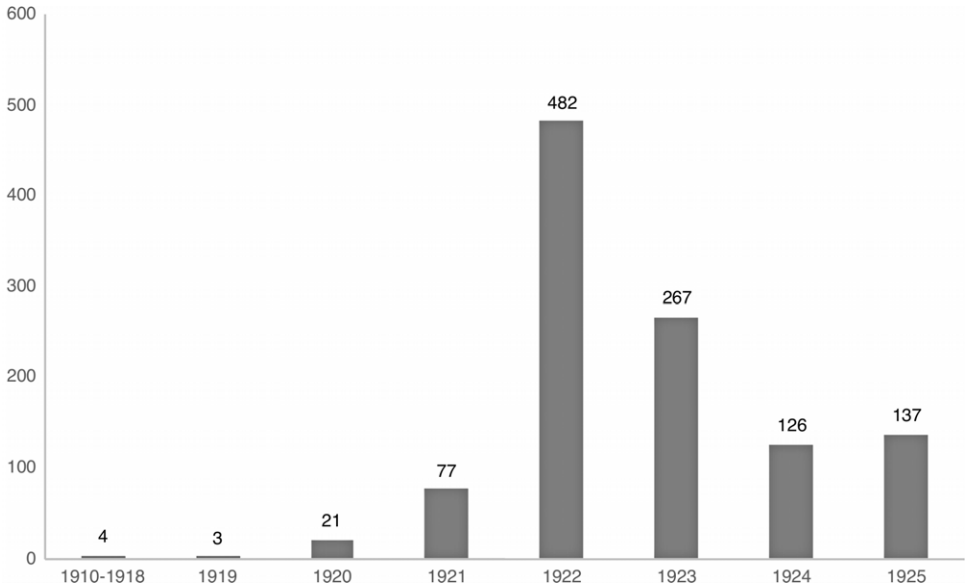

**Figure 2.** Belgian newspaper articles mentioning "Einstein" (1910-1925). Examined titles include: *Le Courrier de l'Escaut, La Dernière Heure, Gazet van Antwerpen, Gazette de Charleroi, De Gentenaar, Het Handelsblad, L'Independance Belge, Journal de Bruxelles, Journal de Charleroi, Het Laatste Nieuws, La Libre Belgique, La Meuse, La Nation Belge, Le Peuple, De$^{ll}$ Schelde, Le Soir, De Standaard, Vers L'Avenir, Le Vingtième Siècle, Het Volk, De Volksgazet, Vooruit,* and *La Wallonie.* These titles are available via the digital infrastructure of the Royal Library in Brussels.

In November 1920, he was introduced as no less than "a glory of science . . . a convinced pacifist, and also Jew, who has been booed by his students to the point of cancelling his course."[21] Apparently, Belgian judgments on Einstein and his theory could differ substantially, also between journalists working for the same newspaper.

Before 1922, Belgian newspaper articles on Einstein thus shared a political framing. Also, they were small in number. The number of Belgian newspaper articles in which Einstein was mentioned increased from a couple in 1919 to well over fifty in 1921. In 1922, however, there were nearly 500 Belgian newspaper articles that mentioned Einstein (figure 2). How did this sudden interest come about?

The key trigger of Belgian interest in relativity and its main inventor was Einstein's visit to France in April 1922. The news that Einstein was about to make a public appearance in France, as the first German scholar in years, was widely covered in the Belgian press. During Einstein's week-long stay in Paris, of which an intellectual confrontation between him and the famous French philosopher Henri Bergson formed the climax (Canales 2015, 3-37), Belgian journalists dedicated dozens of articles to Einstein's personality and work. On a front page article in *Le Peuple*, for example, Einstein's public lecture at the Collège de France was lyrically reviewed, and it was noted with satisfaction that he had received standing ovations.[22] Einstein's nationality remained a topic of concern, but Belgian journalists increasingly gave him the benefit of the doubt. They said he was "first of all a Jew [or] a Swiss," and that he had nothing to do with the "Aryan Germans."[23] Einstein was Einstein for being "*un Allemand exceptionnel*" and for "the splendor of his works."[24] One journalist stated that the anti-German skepticism towards Einstein had largely disappeared: Einstein's theory "had disarmed even the most disheveled patriots. One now exclusively and continuously speaks of him with respect and admiration."[25]

Even though criticism in the French-speaking world did not entirely disappear after 1922, Belgian reporters were right in calling Einstein's public appearance in Paris a success for the rehabilitation of international intellectual cooperation between the Central and the Allied powers. In fact, this had been a pre-intended purpose of Einstein's visit. Initiated and arranged by Einstein's friend and fellow internationalist Paul Langevin, Einstein's trip to Paris was deployed as a means to reestablish relations between French and German scientists (Canales 2015, 17). As an additional effect, Einstein's own reputation among the French improved, because he became less associated with the persistent nationalism of some German scientists (Buchwald et al. 2012, xlvi; Paty 1987, 136). Einstein's visit had a similar effect in Belgium. The sudden fame of Einstein in Belgium in 1922, a reverberation of the massive attention to Einstein in France, was accompanied by a growing recognition and appreciation that Einstein did not share the nationalist agenda of many of his compatriots.

The Belgian orientation on French events is but one illustration of the entanglement between the histories of relativity in France and Belgium. More examples are easily found, as French intermediaries were crucial in bringing Einstein and his theory to the attention of the Belgian public. In fact, the very first Belgian newspaper article about Einstein was a report of an announcement on Eddington's successful expeditions by the vice-president of the Academy of Science in Paris.[26] Moreover, in November 1920, at a moment when Belgian interest in relativity was almost non-existent, the most ardent French advocate of relativity, Paul Langevin, lectured on the topic in Brussels. Two prominent Belgian newspapers reported on Langevin's popular lecture.[27] Given such examples, and because of the strong ties between the francophone French and Belgian intellectual elites in general, one might debate if it is sensible to distinguish separate Belgian and French contexts at all, especially when trying to make sense of a global history of relativity. Indeed, early Belgian attitudes towards relativity were shaped in close connection to French circumstances. A similar dynamic has been described with regards to the early appropriation of Darwinism in Belgium (De Bont 2008, 188).

In this section, I aimed to demonstrate that there were multiple ways in which the global politics of the First World War shaped the local appropriation of relativity in Belgium. First, the war impeded the flow of knowledge across Belgian borders, which I have illustrated by means of the case of De Donder. Second, in the direct aftermath of the war, many Belgians found the German origins of relativity to be problematic, which was part of the reason why the theory was initially ignored. At the same time, however, these circumstances inspired initiatives to promote relativity, such as Lecat's bibliography.

Over the course of the 1920s, the war became less decisive in shaping Belgian perceptions of Einstein and his theory. As Belgian interest in relativity increased, its association with modernism became increasingly significant.

## Relativity and Belgian struggles with modernity

In the summer of 1933, Einstein visited Belgium on his way to the United States (fig. 3). During his stay, he delivered three lectures in Brussels on his own version of the spinor, the "semi-vector" (van Dongen 2004), in front of the elite of Belgian physics. When asked which members of his audience could understand the content of his lectures, Einstein supposedly answered: "Perhaps professor De Donder … Lemaître certainly, and the others I do not think so" (cited in Mawhin 2012, 54-55). The quote suggests that Einstein had a low opinion of the abilities of

---

[25]Ibid.

[26]"Les théories d'Einstein et de Newton," *Gazette de Charleroi*, 11 December 1919.

[27]"Les théories d'Einstein," *Le Peuple,* November 7, 1920; "Conférence de M. Langevin," *Le Soir*, 6 November 1920.

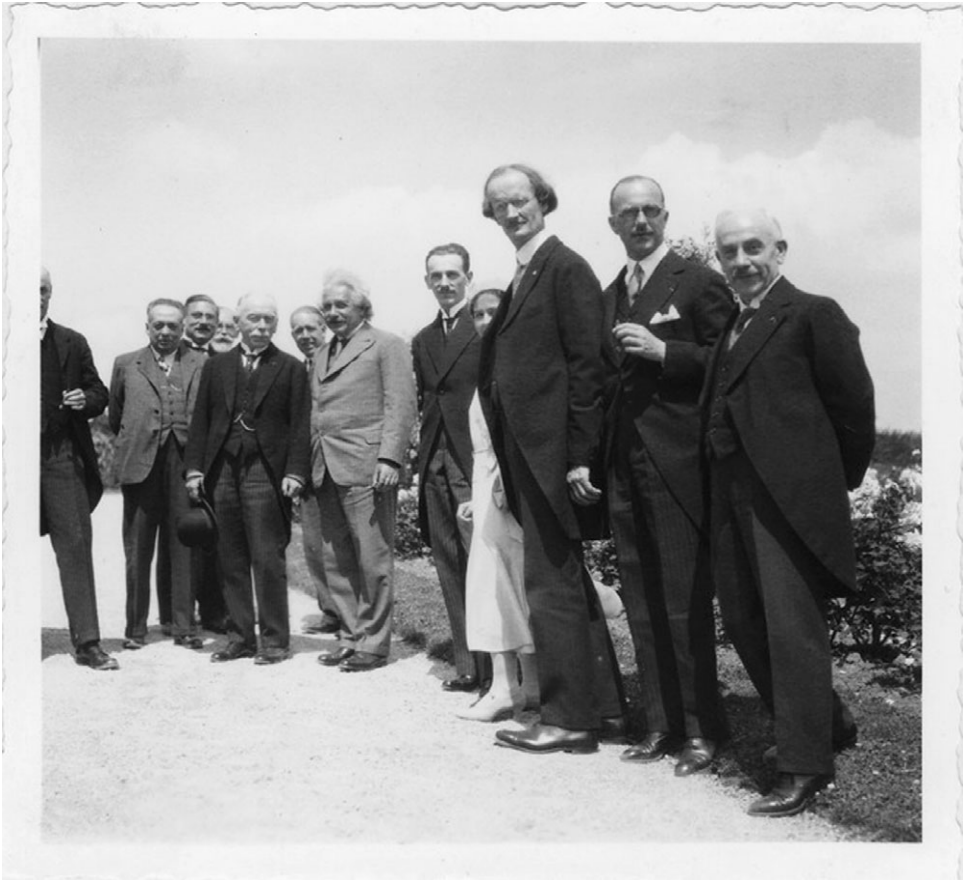

**Figure 3.** Group portrait in Belgium in 1932, made by Belgian King Albert I, including, among other prominent physicists and the Belgian Queen Elisabeth, Théophile De Donder (first from right) and Albert Einstein (sixth from the right). Source: Wikimedia Commons, ETH library.

Belgian physicists. Nonetheless, in the years before, many among them had attempted to contribute to relativity physics. Apart from the considerably well-known Belgian contributions of De Donder and Lemaître, more than twenty other Belgian scholars published on relativity during the first half of the 1920s (see fig. 1, above).

After Einstein's visit to Paris, discussions emerged in Belgium on the meaning and value of the theory of relativity. As elsewhere, its suggested revision of the foundations of space and time urged not only physicists, mathematicians and astronomers, but also philosophers and artists to relate to Einstein's theories. In the discussions that followed, particularly in popular media, relativity was broadly construed as a typical example of modern culture, on which opinions differed substantially. In the Belgian press, parallels between relativity and modern art were common. For example, *La Dernière Heure* introduced the theory and its concepts of a fourth dimension and space curvature as "another modern idea, [suggesting] that 'real' space will eventually appear as some sort of tragic Cubist painting."[28]

---

[28]"Le drame cosmique," *La Dernière Heure*, December 19, 1921. Such parallels between modern art and science, particularly between the kinds involving more than three dimensions, were common at the time. Yet, as Linda Henderson has demonstrated by means of a simple but convincing chronological argument, not relativity theory (which became widely known only in 1919), but late-nineteenth-century developments in non-Euclidean geometry were the main source of inspiration for modern artists to experiment with representing time and space (Henderson 1983).

In Belgian culture, there were different groups that each responded differently to modernization in general, and to the rise of modern science (heralded by the non-classical theory of relativity) in particular. Belgian society around the turn of the twentieth century was "pillarized," meaning that it consisted of distinct communities that fostered distinct worldviews and cultural patterns.[29] This cultural division formed separated cultural-ideological networks and institutions through the whole of the societal domain, including the sciences. The main cultural boundary was between liberals and Catholics.

The liberal-Catholic dichotomy was clearly reflected in the infrastructure of Belgian science (Vanderstraeten 2018, 464-466; Laqua 2013, 85-90). The ULB in Brussels, at which the dominant philosophy was liberal and materialist, found itself ideologically opposed and socially separated from Louvain's *Université Catholique de Louvain* (UCL). The devout Catholic scientist Lecat, who was based in Louvain, at some point announced that he had "refused on several occasions to teach at the fiercely anti-religious University of Brussels" (Lecat 1923, 1). Scientists and science popularizers (or *vulgarisateurs*) based in Louvain and Brussels published in different periodicals and appealed to different audiences. Both camps paid significant attention to relativity in the 1920s, but contrasting views developed on the meaning and implications of the theory.

### Neo-Scholasticism in Louvain

In the late nineteenth century, one of the Vatican's reconciliatory responses to the rise of modern science and the enhanced positivist world view was to aim to revive Thomistic philosophy. This strategy was mainly implemented in Belgium, particularly in Louvain. Pope Leo XIII recognized the UCL as one of the few Catholic universities offering a wide range of scientific training, which made it an excellent location for a global center of Neo-Thomism (or Neo-Scholasticism). In 1889, the Pope approved the foundation of the Higher Insitute of Philosophy in Louvain. It became the most prominent Neo-Scholastic institute, directed by Cardinal Désiré-Joseph Mercier (Vanpaemel 2017, 150-154; Wils 2005, 326-330).

Typical of the philosophy developed at Mercier's institute was the belief that modern science was to be embraced, but only under strict conditions: there could be no conflicts between science and faith (Kragh 2008, 384-385; Hellemans 2001, 121). In order to avoid such conflicts, the science courses at the institute, taught by Catholic scientists such as mathematician Charles de la Vallée-Poussin, physicist and astronomer Ernest Pasquier, and chemist and cosmologist Desiré Nys, were supposed to be free of any "philosophical" content (Vanpaemel 2017, 151). These Neo-Scholastic scholars believed that the scope of modern scientific theories, including relativity, should be curtailed in order to not include philosophical statements.

Beginning in the 1920s, relativity was extensively discussed in journals of the *Société Scientifique de Bruxelles*. This was a society formed by Catholic scientists, aiming at science popularization, with close ties to Louvain's university and Institute for Higher Philosophy. The Scientific Society and its two journals, the professional *Annales de la Société Scientifique de Bruxelles* and the semi-popular *Revue des Questions Scientifiques*, enjoyed a high reputation among Catholics not only in Belgium, but also in France (Nye 1976, 274-275). The many book reviews on relativity published in the *Revue* in the 1920s shared a Neo-Scholastic interpretation of relativity. Most of them were written by Henry Dopp, Jesuit and secretary of the Society. Dopp's attitude toward relativity was not necessarily negative, but he had one major concern. He insisted that Einstein's theory was exclusively significant within the realm of science, and irrelevant to metaphysics or philosophy. In a review of a book by the French physicist Ernest Lémeray, Dopp lamented that "too often ... the author talks about spiritualist and religious convictions as if they employ the very same epistemological methods as scientific theories about the phenomenal world" (Dopp 1922, 451-452). Lémeray had summoned his readers to abandon the classical concept of absolute space. According to Dopp, however, such a decision could not be based on the limited authority of scientific argument.

---

[29]The concept of pillarization (in Dutch: "*verzuiling*") dates from the 1960s (Lijphart 1968). It has also been applied to explain different attitudes towards relativity in the Dutch contect (Klomp 1997).

Another version of the same point was published in a 1923 book by Ghent University's Catholic physics professor Paul Drumaux. Drumaux included an important disclaimer in the conclusion of his otherwise pro-Einsteinian *L'Évidence de la Théorie d'Einstein*: "ask yourself what the physicist introduces at the basis of all of his theories. He introduces numbers. And what does he retrieve? Numbers, nothing but numbers." Drumaux argued that, because "physical essence cannot be caught in numbers … it is inaccessible and therefore impenetrable." Hence, "our knowledge of the Universe can only be … relative" (Drumaux 1923, 69). Dopp and Drumaux's opinions were shared by the Higher Institute's cosmologist Nys, a former student of Wilhelm Ostwald (Kragh 2008, 385). Nys praised the theory of relativity for its mathematical accomplishments, but denied it any ontological significance. In Nys' 1922 *La Notion de l'Espace*, he remarked that the theory of relativity "should be examined from two points of view; one being scientific, the other metaphysical" (Nys 1922, 317). For Nys, the scientific merit of relativity lay in its formulas rather than in its concepts: "the theory of relativity appears to us as a purely mathematical conception, an abstract synthesis, constituted of elements logically chained together by means of artifices of calculation. It would be reckless to regard the theory as a representation of reality" (ibid., 321).

So, these Belgian Catholic scientists and philosophers felt it was needed (and possible) to deny and remove any metaphysical connotations and interpretations from the theory of relativity.[30] While they generally acknowledged the value of the theory for scientific issues, they denied it authority beyond the immediate realm of mathematical science. Similar views resonated in popular Catholic media. A front-page article in *Het Volk* posited that relativity had nothing to say about philosophical and religious questions such as the nature of space and time. Popular advocates of the theory of relativity were blamed for concealing "that they were discussing purely mathematical thought," and for applying the theory of relativity to "non-scientific" matters.[31]

Someone Belgian Catholics could accuse of such practices was De Donder. How did his relation to Einstein's theory develop after the war? And what was his position within the divided Belgian culture of science?

### De Donder's school in Brussels

After the war, De Donder's scientific status increased, both in Belgium and abroad. He formed a research group at the ULB that focused on general relativity – including Henri Vanderlinden, Maurice Nuyens, Frans van den Dungen, and Carlo de Jans – which published numerous articles on the topic in francophone journals. These articles often built on previous, mathematically-oriented work by De Donder. In 1921, De Donder published the book *La Gravifique Einsteinienne*, which included most of his earlier contributions to general relativity (De Donder 1921). By now, De Donder had managed to clarify the relation of his work to Hilbert, and fixed the error in his variational derivation of the field equations.[32] De Donder's *Gravifique* received considerable international attention, in Britain and Italy for example, due to favorable opinions expressed by Eddington and Tullio Levi-Civita (Bosquet 1987, 237; Cattani 2010, 137-143).

---

[30] In the early twentieth century, diverging approaches were employed to reconcile the domains of religion, philosophy, and science, including relativity. The approaches of some, including Arthur Eddington, were more synthetic than those of others, for example the Belgian Neo-Scholastics (cf. Graham 1982).

[31] "De wonder-geleerde EINSTEIN…," *Het Volk*, 24 October 1922.

[32] Only a year before publication, however, De Donder had made another attempt to convince Einstein of the advantages of his alternative formal approach to general relativity (De Donder to Einstein, 3 August 1920, in Buchwald et al. 2006a, 363-364; trans. in Buchwald et al. 2006b, 227). But when Einstein replied that "I cannot make any clear sense of your equations" (Einstein to De Donder, 11 August 1920, in Buchwald et al. 2006a, 370-371; trans. in Buchwald et al. 2006b, 231-232), De Donder conceded and announced that he had decided to "transcribe all my results in your notation" (De Donder to Einstein, 18 August 1920, in Buchwald et al. 2006a, 376-378; trans.in Buchwald et al. 2006b, 236-237).

De Donder's post-war publications had their greatest influence in France, however. In 1924, an editor of the renowned French science publisher Gauthier-Villars praised him as "one of the specialists among specialists" in relativity.[33] This status was well-deserved, since in earlier years De Donder had repeatedly published in journals issued by the French Academy of Science and attended multiple French meetings and conferences on relativity theory, including some during Einstein's visit to France (Bosquet 1987, 237). In Belgian media, in turn, De Donder regularly reviewed books by French relativists, including the book by Lémeray which Dopp had criticized (De Donder 1922a; De Donder 1922b). In Germany, however, De Donder's influence remained limited. His three German correspondents (including Einstein) stood in sharp contrast to 22 from France, and more than 70 international correspondents in total (Glansdorff et al. 1987, 30).

The particular international orientation of De Donder and his students indicates the lasting influence of the First World War on the scientific networks through which the work of relativists circulated during the 1920s. As Catherine Goldstein and Jim Ritter have shown, relativists from Allied and Central power countries often worked on the very same subjects, yet remained in distinct communities (Goldstein and Ritter 2003, 101). This did not change when, from the mid-1920s onwards, De Donder and his students shifted their attention from general relativity towards quantum theory, like most of their international colleagues (Blum et al. 2015). Helge Kragh has noted that while their work was mathematically solid, it fell outside of the mainstream of quantum physics, where it made little impact. According to Kragh, De Donder and his students "remained isolated from the trend in quantum physics which took place outside the French-speaking world" (Kragh 1984, 1031).

De Donder's views on the philosophical importance of relativity were diametrically opposed to those of the Belgian Neo-Scholastics. In one of his first lectures as professor in Brussels, De Donder claimed that "mathematical physics contains the purest picture of nature that might spawn in the human mind" (cited in Glansdorff et al. 1987, 18). Therefore, he held that the theory of relativity was philosophically relevant, contrary to what his Catholic compatriots claimed. In the liberal journal *Le Flambeau,* De Donder remarked that some had erroneously claimed "that these theories are only mathematics, and that they do not provide any explication of physical realities!" (De Donder 1920b, 715). According to De Donder, those who denied the philosophical significance of Einstein's theory had just failed to grasp the mathematics involved. He concluded by stressing the ontological significance of the theory of relativity: "the absolute is the object of general relativity" (De Donder 1920b, 731). After publishing this article, he proudly wrote Einstein to announce that he had "attempted to present the handsome philosophy that emerges from your theories" for a general audience.[34]

De Donder's opposition to Neo-Scholastic ideas comes as no surprise when considering the liberal background of his alma mater and long-time employer, the ULB. The divide between the school of De Donder and Catholic scientists from Louvain was both ideological and institutional. De Donder and his students published in journals of the Royal Academy, and sometimes abroad, but never in Catholic journals. Catholic scientists, in turn, rarely published at the Royal Academy.[35] These findings reinforce existing images of Belgian science and society as divided into two different "pillars" at the beginning of the twentieth century.

Yet this image deserves nuance, as crucial cross-links between the Catholic and the liberal pillars were established precisely due to the introduction of relativity to Belgian science.[36] Strikingly, priest-cosmologist Georges Lemaître was introduced to the theory of relativity in Louvain, after which he further pursued his studies under the wings of De Donder. Lemaître's example suggests that the ideological boundaries in Belgian science were becoming less strict during the inter-war period. What else do Lemaître's first introductions to relativistic physics reveal about the history of relativity in Belgium? And, vice versa, what can the Belgian context teach us about Lemaître's introduction to general relativity?

---

[33]Henri Villat to De Donder, 30 October 1924, ULB De Donder Archives.

[34]De Donder to Einstein, 3 August 1920, in Buchwald et al. 2006a, 363-364; trans. in Buchwald et al. 2006b, 227.

[35]In total, around 25 articles on relativity appeared at the *Bulletin* and *Mémoires* of the Royal Academy between 1920 and 1925, only two of which were not directly linked to either De Donder or one of his students.

[36]Pasquier to De Donder, 17 December 1922, ULB De Donder Archives.

## Lemaître's cosmology and Belgian relativity

Georges Lemaître proposed one of the first expanding and relativistic models of the universe in 1927, and the first Big Bang model of the universe in 1931 (Kragh 1999, 22-79). The results of this study shed light on the historical background of these theories. Indeed, Lemaître's familiarization with relativity and relativistic cosmology was crucially dependent on the particular ways in which relativity had been appropriated in the Belgian context.

Lemaître obtained a doctoral degree in mathematics at the Catholic University of Louvain in 1920 and was ordained as a priest in 1923, after attending the Saint-Rombaut seminary in the city of Mechelen in the years in between. In Belgium, such a hybrid scientific-religious career was not uncommon, which underlines the complex dynamics between science and religion in the country, particularly in Louvain. Lemaître invariably demarcated scientific from religious truth, as did most Belgian Neo-Scholastics. Typically, Lemaître's academic training was Belgian, which had been predetermined by the kingdom's strong orientation on engineering. In 1911, he enrolled in the program of mining engineering taught at the UCL. Subsequently, Lemaître's academic studies were interrupted by the outbreak of World War I. Lemaître was a proud patriot, and fought at the front lines. Only after he had returned from war and abandoned the idea of becoming an engineer, Lemaître shifted towards mathematics and physics, and specialized in Thomistic philosophy at the Higher Institute for Philosophy (Lambert 2015, 53-66). Given the above, it is fair to establish that Lemaître's early academic career unfolded in typically Belgian circumstances.

The same goes for his introduction to relativity. Lemaître was introduced to relativity in Louvain. During his mathematical studies at the UCL, canon René de Muynck encouraged Lemaître to focus on physics (Vanpaemel 2017, 124), and Nys introduced him to cosmology (Kragh and Lambert 2007, 459). Furthermore, Lemaître became familiar with the international literature on relativity when he read book reviews in the *Revue des Questions Scientifiques* (Lambert 2015, 67-85). During his time spent at the seminary, Lemaître was granted permission by Cardinal Mercier to further pursue his studies of relativity, and to consult relevant literature at the Royal Library in Brussels (Lambert 2015). Noting this Belgian-Catholic background, it is comprehensible why Lemaître's seminal 1927 paper on the expanding universe was published in the main Catholic scientific journal of Belgium: the Scientific Society's *Annales de la Société Scientifique de Bruxelles* (Lemaître 1927). In the francophone Catholic community, this journal was prominent. Outside these circles, however, the *Annales* was obscure, which helps to explain why most relativists and cosmologists initially overlooked Lemaître's contribution.[37]

It would be too simplistic to interpret Lemaître's first encounters with relativity as only shaped by his Catholicism. Lemaître was on good terms with the liberal De Donder, and corresponded about the latter's contributions to general relativity.[38] Indeed, De Donder's formal approach to relativity strongly influenced Lemaître. In *La Physique d'Einstein,* a long and detailed, yet unpublished manuscript about relativity dated 1922, Lemaître engaged with the notation used by

---

[37]Jean Eisenstaedt has expressed his amazement at the lack of attention the article received and suggested it might have been due to the circumstance it had been written in French. Not just the French background of the *Annales* was crucial, however, but also its *Catholic* imprint (Eisenstaedt 1993, 378-379n44).

[38]In one letter, De Donder quoted passages from texts by Lorentz and Eddington in which his own contributions were acknowledged. De Donder to Lemaître, 10 October 1923, ULB De Donder Archives.

De Donder in *La Gravifique Einsteinienne* (Lemaître [1922] 1996; [1922] 2019). In his first publication of 1923, Lemaître even immersed himself in De Donder's variational calculus (Lambert 2012, 10). Furthermore, De Donder was on a jury that awarded Lemaître a fellowship to visit Eddington in Cambridge. Lemaître's visit to Cambridge in 1923 and 1924 was a considerable success (Lambert 2015, 79-83). After Lemaître left Cambridge, Eddington wrote to De Donder to inform him that "I found M. Le Maître (sic) a very brilliant student, wonderfully quick and clear-sighted, and of great mathematical ability."[39]

Especially from the 1930s onwards, Lemaître and his work gained an outstanding reputation. This was partly due to Eddington, who not only welcomed Lemaître as a student, but also published Lemaître's English translation of his 1927 *Annales* paper in 1931 (Eisenstaedt 1993, 361-362; Kragh 1999, 31-36; Livio 2011). The translation made it possible for Lemaître's work to circulate within a transnational community of relativists and cosmologists. But before Eddington's intervention, Lemaître's 1927 paper had been largely neglected, even by Eddington himself. This once more illustrates the isolated position of early-twentieth-century Belgian relativists such as De Donder and Lemaître.

Lemaître's eventual success, however, shows that Belgian scientists did increasingly manage to acquire positions within relevant transnational scientific communities. Key to Lemaître's eventual visibility and success was that he had managed to transcend the ideological divisions within Belgian scientific culture. This reflects larger trends in Belgian science and society, which became less ideologically divided and more internationally relevant during the interwar period. This is also evident, for example, from the institutional development of the sciences in Louvain. As Geert Vanpaemel has argued, Louvain's faculty of science developed "from a Catholic institute for education in a pillarized society into an open research environment embedded within the international scientific community" (Vanpaemel 2017, 167).

## Conclusion: Transnational flows and global patterns in the history of relativity

By studying the different ways in which relativity was appropriated in Belgium, I have added yet another title to the already long list of nationally focused case studies. But I have also argued that the history of relativity should not remain confined to local perspectives. What kind of transnational picture emerges when this case study is brought into contact with others? And how should this be done?

Comparison may provide a starting point of transnational historiography. However, as James Secord has noted, comparative work "can all too often end up reaffirming national boundaries" (Secord 2004, 669). Comparative studies in the history of relativity in particular have focused on the differences across national borders rather than on shared patterns. Hence, it is important to consider using other methods as well. The editors of the *Palgrave Dictionary of Transnational Historiography* have argued that historians wishing to transcend national boundaries ought to pay attention to "links and flows," and track "people, ideas, products, processes and patterns that operate over, across through, beyond, above, under, or in-between polities and societies" (Iriye and Saunier 2009, xxiii; see also Turchetti et al. 2012, 327). What can we learn about the global history of relativity, when we follow up on this suggestion and trace specific flows of manuscripts, scientists, and letters across national borders, and thereby expose the transnational scientific networks and institutions that regulated these flows?

### *Theoretical physics and marginalization*

The fact that relativity was seen as a signature expression of the novel subdiscipline of theoretical physics proves essential for understanding public and scientific reactions transnationally. I have argued that the absence of a theoretically minded tradition in physics in Belgium explains the initial lack of interest in Einstein's work (leaving aside De Donder's isolated attempts). The relatively late appropriation of relativity in Spain has been explained similarly (Glick 2014). The practice of theoretical physics had not acquired a place within local scientific traditions in Belgium, nor in Spain, when relativity became prominent internationally. As a result, there was no context in which relativity could be appropriated.

[39]Eddington to De Donder, 24 December 1924 (cited in Bosquet 1987, 250).

At places where theoretical physics was practiced, this often led to clashes between locally dominant research programs and the Einsteinian approach. For instance, British physicists did not regard Einstein's theory as particularly useful compared to their own electronic theory of matter (Warwick 1992; Warwick 2003). And although Dutch physicists were generally ready to embrace Einstein's theories, Lorentz never really became convinced of the necessity to abandon the concept of ether (Kox 1988). In these cases, there was no passive "reception" of Einstein's concepts and theoretical practice. Rather, physicists created alternative, local versions of relativity, sometimes in consultation with Einstein (Renn 2007, 1-2). As noted above, I propose to call this internationally recurring process "appropriation" or "co-creation."

Such co-creation also typifies the case of De Donder, who, on his own, provided the Belgian scientific context in which relativity was initially appropriated. During his introduction to relativity physics, De Donder chose to adhere to his own mathematical methods, creating his own version of general relativity. Although he eventually conformed to the Einsteinian program, which became the international standard, his first reaction was to challenge the approach of his German colleague. De Donder made a priority claim, and called into question the validity of Einstein's methods.

Thus, when seen together, the different case studies make clear that the appropriation of relativity into various local contexts was an active process, during which the form and meaning of the theory could change considerably. Simultaneously, local styles of scientific practice were transformed. These processes can be better understood from a transnational perspective, since, as John Krige has argued, "the construction of a transnational knowledge community collides with vested local interests and their histories that resist its pressure to homogenize and to standardize" (Krige 2019b, 16).

The internationally differing form and prestige of theoretical physics can be linked to the transnational history of relativity in yet another way. Milena Wazeck has argued that the main source of opposition against Einstein and his theory in the 1920s was the marginalization of certain groups, including amateur scientists or "world riddle solvers," philosophers, and in particular experimental physicists such as Ernst Gehrcke and Philipp Lenard, by the rise of the highly mathematized and professionalized sub-discipline of theoretical physics (Wazeck 2014). Wazeck's argument is convincing, but mainly based on the study of specific networks of anti-relativists, involving mostly Germans and Americans. Can her framework also be applied beyond these networks, as to include other national contexts?

In Belgium, the situation was different from that in Germany where, as theoretical physics enjoyed little prestige, and was not institutionalized. Yet, this may actually explain the lack of structural opposition to relativity among Belgian scientists. Unlike their German colleagues, Belgian experimentalists doing original research, like Pierre de Heen, Edmond van Aubel, and Verschaffelt, were not threatened by the rivalling approach of theoretical physics that was symbolized by the theory of relativity. Thus, Wazeck's point of view aids in understanding the Belgian case, increasing the plausibility of marginalization as a transnationally valid perspective.

### *Transnational networks and the First World War*

Yet, there was resistance in Belgium. This suggests that a framework focusing exclusively on scientific marginalization may not suffice to capture the global history of relativity. Indeed, politics mattered as well. The findings of this study reinforce Matthew Stanley's assertion that "questions of politics . . . molded not only the creation of relativity but the way the world came to first meet Einstein" (Stanley 2019, Prologue). This brings me to another global pattern in the history of relativity: the omnipresence of the First World War. In many studied countries, the First World War shaped nationalist and internationalist discourse on Einstein and his theory. It is ironic that while some of his compatriots blamed him for not being German enough, in other countries Einstein faced the problem of being dismissed precisely for being German. In Germany, Einstein's anti-militaristic, democratic stance provoked hostile responses of nationalists during the post-war years (van Dongen 2007). In France and Britain, the prevalence of anti-German sentiment after the war implied that Einstein ran the risk of being primarily perceived, and subsequently discarded, as an exponent of German science, despite the fact that he did not sign the manifesto of 93 (Stanley 2003, 58).

At first glance, the way in which the First World War staged germanophobic discourse in Belgium strongly reminds one of the French case; in both Belgium and France, Einstein's relation to Germany was the most prominent topic of concern in the press (Van Kimmenade 2010). But there is a striking parallel between Belgium and the Netherlands as well. In both of the Low Countries, internationalists saw Einstein as a kindred spirit, and promoted his theory (van Besouw and van Dongen 2013). Likewise, Eddington's effort to test general relativity was motivated by an internationalist agenda, that is, by the hope that such an international expedition would repair the damaged intellectual relations and "heal the wounds of war" (Stanley 2003; Kennefick 2019). Eddington, indeed, "realized that relativity could be the key to restoring" international scientific networks (Stanley 2019, Prologue). So, not only in Belgium but across Europe, internationalism was a factor that promoted the scientific and public prestige of Einstein and his theory.

The Belgian case also illustrates how the networks through which relativity physics circulated were shaped by the First World War. Lorentz's central position within these networks was particularly striking. The pivotal position of Dutch physicists in the transnational circulation of relativity has been remarked upon before. Around World War I, they enabled the flow of goods (e.g., manuscripts and publications) and people (e.g., Einstein himself) through international communities of relativists. For example, Eddington became familiar with Einstein's work and approach via De Sitter (Warwick 2003, 447), and Paul Ehrenfest operated as a mediator between Dutch and Russian communities of physicists (Vizgin and Gorelik 1987). In addition, the case of Belgium has made clear that, after the war, separate communities of relativists formed along Central and Allied lines. Accordingly, De Donder's works were read in the French- and English-speaking world, but not in Germany. During the 1920s, De Donder and his students acquired an increasingly prominent position within the transnational, Allied network of relativists. De Donder reviewed French books on relativity in Belgian periodicals, and in the French academic world he was even regarded as one of the specialists in the field.

Here I have only been able to carve out a small part of the transnational networks that played a role in the global circulation of relativity. Surely, a further deepening of the transnational scientific networks that formed around the war has the potential of yielding a clearer picture of how relativity gained global prestige, and how it sustained it over the course of the twentieth century (see Lalli 2017).

### Revolutionary contexts

Another emerging global pattern relates to the fact that interest in the theory of relativity developed during a period of political revolutions that were taking place throughout Europe. Simultaneously, the arts were revolutionized. These revolutionary conditions shaped how Einstein and his theory were perceived globally. As a pacifist, democrat, and Jew, Einstein was considered by many to represent revolutionary political thought. His mathematically advanced theory was seen as the main exponent of modern theoretical physics, and the revolutionary philosophical foundations of his theory became associated with the non-classical foundations of modern art (Henderson 1983). For political, cultural, and scientific conservatives, consequently, the interest in relativity reminded them of revolutions in the scientific, cultural and political domain (van Dongen 2012). Relativity and Einstein were generally identified as revolutionary heralds of modernization, and this invited both enthusiasm and reactionary backlash.

In Belgium, reactionary objections were mostly raised by Catholics. The influence of Catholic thought was not confined to the Belgian context, however. Similar ideologically colored responses emerged among Dutch Catholics (Klomp 1997, 102-109). Thomas Glick, studying the Spanish context, encountered similar dynamics. However, he holds that some conservative Catholics actually found the philosophical appeal of relativity theory attractive, for they preferred it over the positivist worldview that they associated with Newtonianism and Kantianism (Glick 1987a, 238-243). The influence of Catholicism on global attitudes towards relativity yet deserves a more systematic, transnational approach. For one thing, such an approach could shed light on the opposition to relativity of the famous French anti-relativist Pierre Duhem, who was a devout Catholic and occasionally published in the Belgian *Revue des Questions Scientifiques* (Kragh 2008, 380).

Einstein and relativity's "revolutionary" reputation proved relevant for their perception across Europe. Italian advocates of relativity initially portrayed relativity as a revolutionary theory. But when the Italian political climate changed and became less revolution-minded, they responded by rendering it as a theory which brought only gradual change (Reeves 1987). In the Netherlands, the theory of relativity became an ingredient in heated, society-wide debates on the status of modern culture. Because of its alleged incomprehensibility and counter-intuitiveness, it was associated with avant-gardism (van Besouw and van Dongen 2013, 96; Baneke 2008, 154-162). In Germany, Einstein's name was smeared as an expression of Dada culture and avant-gardism. The theory of relativity represented a modernist spirit that many Germans conservatives found worthy of condemnation. One of Einstein's loudest German opponents, Paul Weyland, argued that the mass interest in relativity must be understood as a sign of a "scientific Dadaism" which had taken over German academia (Van Dongen 2007, 216).

The modernist, revolutionary reputation of Einstein and his theory may assist in understanding dynamics surrounding the theory beyond the Western World as well. Danian Hu has argued that Chinese attitudes towards Einstein and relativity depended on shifting cultural and political identities over the course of the twentieth century. Initial openness and enthusiasm for relativity, which originated from Japanese missionary work, went hand in hand with the Chinese assimilation of Western scientific ideals at the beginning of the twentieth century. Chinese intellectuals eagerly embraced relativity's revolutionary nature, as they were "thirsty for revolution – both political and scientific" (Hu 2007, 543). After the Second World War, however, the close association of relativity with Western culture triggered a Chinese anti-relativity campaign. Communist critiques of relativity emerged in the 1950s, and during the Cultural Revolution (1966-1976) it was widely disapproved (Hu 2007, 551).

The Chinese case exemplifies that Einstein and relativity were considered as symbols of Western cultural modernization. Relativity's modernist reputation thus proves relevant to understanding attitudes toward the theory not only in the Western World, but globally. This also becomes clear from Einstein's widely reported visits to South America and Asia in the 1920s. In Japan, Argentina, and Brazil, Einstein and relativity were treated as forerunners of modernization (Glick 1987b, 389). Furthermore, the Chinese case illustrates the function of transnational scientific networks in the global circulation of relativity, and the possibility of transnationally oriented histories in tracing these networks: Hu shows that relativity physics was introduced to Chinese scientists through Japanese intermediaries, which leads him to identify Japan as a "hub of scientific transmission" in East-Asia (Hu 2007, 541-542).

It is important to note that the different patterns identified above cannot be easily drawn apart. Indeed, it would be too simplistic to reduce the dynamic of the global history of relativity in separate scientific, cultural, or political realms, as the boundaries between these patterns cannot be drawn. For example, the Belgian Catholic struggle with modernity was not confined to the scientific domain, it was fought in broader political-cultural realms as well. And scientific conservatives in Germany, such as Ernst Gehrcke and Philip Lenard, often felt threatened by cultural and political revolutions as well, which they understood as being part of the same revolutionary spirit haunting Western European society during the interwar period. In Weimar Germany, as in post-war Belgium and other Western European countries, politics and science became strongly intertwined (Rowe 2012). As Jeroen van Dongen has argued, "some of the most visible anti-relativists also feared marginalization of their social and political positions and conservative cultural values" (van Dongen 2012, 167; see also van Dongen 2010).

Ultimately, then, international opposition to Einstein and his theory can be understood as responses to processes of marginalization, comprising scientific, cultural and political aspects. Likewise, endorsement of Einstein and his theory especially developed in circles that identified with the revolutionary scientific, cultural, and political spirit of relativity and its main protagonist. Specific manifestations of this underlying dynamics materialized in local contexts such as Brussels and Louvain, but its general features recurred globally (Van Besouw and Van Dongen 2013, 103).

**Acknowledgments.** First of all, I thank two anonymous reviewers for helpful comments. As for other sources of help and inspiration, I have presented earlier versions of this article at "Science and the First World War: The Aftermath" at the Royal Society in London in September 2018, "100 Years of Applying and Interpreting General Relativity" at the University of Bern in September 2017, and the First Teylers Meeting on History of Astronomy and Physics, at Teylers Museum, Haarlem in April 2017. I thank my audiences for stimulating discussions. Furthermore, I thank Jip van Besouw, Roberto Lalli, Emma Mojet, Chaokang Tai, and Daan Wegener for comments on earlier drafts of this article, and Geert Vanpaemel and Raf de Bont for invaluable advice. Finally, I am much indebted to Jeroen van Dongen, who introduced me to this topic when I was a student in Utrecht's History and Philosophy of Science master's program, and since then made numerous contributions to my research.

**Sjang ten Hagen** is a historian of the sciences and humanities, who is interested in the sharing of knowledge across disciplinary and geographical boundaries. He defended his dissertation – entitled "History and Physics Entangled: Disciplinary Intersections in the Long Nineteenth Century" – at the University of Amsterdam in 2021. He is currently a postdoctoral researcher at Leiden University, the Netherlands.